\documentclass[iicol, pdflatex, sn-basic]{sn-jnl}

\usepackage{commath}
\usepackage{comment}
\usepackage{lineno}
\usepackage{graphicx}
\usepackage{multirow}%
\usepackage{amsmath,amssymb,amsfonts}%
\usepackage{amsthm}%
\usepackage{mathrsfs}%
\usepackage[title]{appendix}%
\usepackage{xcolor}%
\usepackage{textcomp}%
\usepackage{manyfoot}%
\usepackage{booktabs}%
\usepackage{algorithm}%
\usepackage{algorithmicx}%
\usepackage{algpseudocode}%
\usepackage{listings}%

\begin{document}

\title{Polarization from Rapidly Rotating Massive Stars}

\author*[1]{J. Patrick Harrington}
\affil[1]{Department of Astronomy, University of Maryland, College Park, MD, USA}

\author*[2]{Richard Ignace}
\affil[2]{Department of Physics \& Astronomy,
East Tennessee State University,
Johnson City, TN 37614, USA}

\author*[3]{Kenneth G.\ Gayley}
\affil[3]{Department of Physics \& Astronomy, University of Iowa, Iowa City, IA, USA}

\abstract{
Stellar rotation has long been recognized as important to the evolution of stars, by virtue of the chemical mixing it can induce and how it interacts with binary mass transfer.  Binary interaction and rapid rotation are both common in massive stars and involve processes of angular momentum distribution and transport. An important question is how this angular momentum transport leads to the creation of two important classes of rapidly rotating massive stars, Be stars defined by disklike emission lines, and Bn stars defined by rotationally broadened absorption lines.  A related question is what limits this rotation places on how conservative the mass transfer can be. Central to addressing these issues is knowledge of how close to rotational break-up stars can get before they produce a disk.  Here we calculate diagnostics of this rotational criticality using the continuum polarization arising from a combination of rotational stellar distortion (i.e., oblateness) and redistribution of stellar flux (i.e., gravity darkening), and compare polarizations produced in the von Zeipel approximation with the approach of Espinosa Lara \& Rieutord (ELR).  Both produce similar photospheric polarizations that rise significantly in the far ultraviolet (FUV) for B stars, with a stronger signal in the von Zeipel case.  For early main-sequence and subgiant stars, it reaches a maximum of $\sim 1$\% at 140 nm for stars rotating at 98\% of critical, when seen edge-on.  Rotational rates above 80\% critical result in polarizations of several tenths of a percent, at high inclination.  Even at a low inclination of $i=40^\circ$, models at 98\% critical show polarization in excess of 0.1\% down to 200~nm.  These predicted stable signal strengths indicate that determinations of near-critical rotations in B stars could be achieved with future spectropolarimetric instrumentation that can reach deep into the FUV, such as CASSTOR, the {\em Polstar} mission concept, or the POLLUX detector design.
}




\maketitle

\section{Introduction}

The influence of stellar rotation on the appearance of stars and their evolution has a long and storied history \citep[e.g.,][]{1929MNRAS..90...54E, 1950MNRAS.110..548S, 1967ApJ...150..571G, 1978trs..book.....T, 1984ApJ...279..763N, 1992A&A...265..115Z, 2000ARA&A..38..143M, 2002A&A...381..923S, 2012A&A...537A.146E}. Several pieces of observational evidence have brought the issue of stellar rotation into particular prominence for massive stars at this time.

Massive stars are now known commonly to be born as binaries and multiples \citep{2012Sci...337..444S}.  A substantial fraction of massive stars can experience mass transfer during their evolution \citep{2013ApJ...764..166D}.  Massive stars tend to be fairly fast rotators at the level of 100~km/s and greater \citep{2010ApJ...722..605H}. At such speeds rotation can induce rotational mixing.   Two important subtypes exhibiting higher levels of rotation are the emission-line Be stars and the broad-line Bn stars \citep{2017MNRAS.465.4795B, 2020A&A...634A..18C}.  These stars are common enough to make them valuable laboratories for understanding massive-star rotation, and although Be stars are often studied to understand why they have produced disks, it is equally important to juxtapose them against Bn stars to understand why the latter have not (yet) done so.

Such a study requires quantifying the rotation rates of the stars, yet currently the uncertainties remain large \citep[e.g.,][]{2004MNRAS.350..189T}.  Some nearby examples have been observed with optical interferometry and appear consistent with near-critical rotation at levels of 90\% breakup and more \citep[e.g.,][]{2003A&A...407L..47D, 2005ApJ...628..439M, 2006ApJ...645..664A, 2012A&ARv..20...51V}.  Both terms ``critical'' and ``breakup" refer to when the rotational speed at the equator causes the effective gravity to reach zero there, such that a rigidly rotating equator would be in orbit.  

Given that Be stars have Keplerian disks during main-sequence and post-main-sequence stages of evolution \citep{2001PASJ...53..119O, 2013A&ARv..21...69R} is challenging to understand unless the stars are nearly critical rotators.  Additionally, the number of detected and candidate stripped-core OB subdwarf companions of Be stars is steadily growing \citep{2019ApJ...885..147K, 2024ApJ...962...70K}.  In such systems, mass transfer provides a ready source of orbital angular momentum for spinning up the Be stars \citep{2025arXiv250514780L}, suggesting this may be a typical pathway.

There are numerous studies for Be and Bn stars populations in an attempt to understand better the factors that lead to rapid rotation, and its consequences.  In the Galaxy nearly 20\% of B stars display the Be phenomenon, although its prevalence is more common among earlier spectral types than later ones \citep[e.g.,][]{2003PASP..115.1153P}.  For example, around a third of stars among the B1 type are Be.  This is a substantial fraction indicating that fast rotation can be common.  Indeed it may be an underestimate since some Be stars are known to lose and regenerate their disks, so stars can move in and out of the Be classification \citep[e.g., see discussion in][]{2013A&ARv..21...69R}.  In the lower metallicity LMC, the Be phenomenon is even more prevalent \citep{1999A&A...346..459M, 2006ApJ...652..458W}.

As noted, there is a theoretical driver to suspect that Be stars are rotating at near critical speeds.  This is because of the need to impart angular momentum to gas at the stellar atmosphere in order for it to become part of the viscous Keplerian disk \citep[e.g.,][]{2013A&A...553A..25G}.  One mechanism is for pulsational modes that conspire to launch material off the star into orbit with some fraction of the gas gaining angular momentum while the rest falls back to the atmosphere.  Evidence for a pulsational connection comes from studies such as \cite{2003A&A...411..229R}.  Recent work by \cite{2025A&A...699A..82L} shows examples of blob material injected fron the star into low orbit which subsequently disperses to join the disk.  

However, the gain in rotation speed for this material is at the level of several percent of critical speed, perhaps 10\%.  This would require rapid rotation of the equator at the level of 90\% critical.  And yet, \cite{2005ApJ...634..585C} has suggested that the $v\sin i$ distributions for Be stars are consistent with a range of rotation rates that go much lower. While indeed some fraction of Be stars are consistent with near-critical rotation, about half are suggested to rotate below 70\% of breakup.  Although still quite fast, the pulsational mechanism for imparting angular momentum to the gas so that it enters into disk orbit appears inadequate by a significant factor.  Consequently, it remains unclear how to source the disks of Be stars, if indeed the stars can be rigid rotators with significantly subcritical equatorial rotation.

Linear polarimetry is capable of providing an observational signature of how close stars rotate with respect to the critical rate.  While spatially unresolved spherically symmetric stars yield complete polarization cancellation, fast rotation serves to break spherical symmetry in two important ways.  First, the star becomes rotationally distorted, or oblate, which by itself results in a net polarization of the starlight.  Second, the temperature distribution becomes a function of latitude \citep{1924MNRAS..84..684V, 2011A&A...533A..43E}, exhibiting what is termed gravity darkening at the equator.  

This symmetry breaking and consequent implications for continuum polarization have long been known and described through radiative transfer calculations by \cite{1968ApJ...151.1051H, 1991ApJS...77..541C, 1991MNRAS.253..167C}. More recent calculations have led to applications to fast rotating massive hot stars that include Regulus \citep{2017NatAs...1..690C}, $\zeta$ Pup \citep{2024MNRAS.529..374B}, $\epsilon$ Sgr \citep{2024ApJ...972..103B}, and $\zeta$ Aql \citep{2023MNRAS.520.1193H}.

One important result of these models and their applications to data is that relatively large polarizations for hot stars are mainly achieved at far-ultraviolet (FUV) wavelengths, whereas polarization at optical wavelengths is quite low.  In addition, there is new impetus to revisit the issue from two separate directions.  On the observational side, there are a number of science drivers in support of ultraviolet (UV) spectropolarimetry from spaceborne telescopes \citep{2023MmSAI..94b.294N, 2024BSRSL..93..156I, 2024sf2a.conf..457G}.  Examples include the {\em Arago} mission concept \citep{2023SPIE12777E..4FM}, the {\em Polstar} mission concept \citep{2025AAS...24535317D}, CASSTOR \citep{2025arXiv250703956N}, and the POLLUX instrument \citep{2024SPIE13093E..3VG}.  On the theoretical side, previous calculations for the continuum polarization of massive stars as a function of rotation rate have typically employed the von Zeipel theorem for the temperature distribution with local gravity \citep{1924MNRAS..84..684V, 1999A&A...347..185M}.  However, \cite{2011A&A...533A..43E} have presented an improved treatment for the temperature distribution which differs from von Zeipel, but the larger implications for the Be/Bn stellar population remains to be explored.

The combination of seeking a robust diagnostic of massive star rotation rate, of prospects for future UV spectropolarimetry capabilities, and of advances in our understanding of stellar atmosphere properties at near-critical rotations have motivated us to present new calculations of short-wavelength continuum polarization for rapidly rotating hot stars.  In Section 2 the von Zeipel theorem is reviewed. Also, expressions based on the latitude-dependent effective gravity are used to derive the shape of the star assuming a Roche potential. Section 3 then compares the temperature distributions from von Zeipel with ELR and shows that continuum polarizations from fast rotationg stars are consistently somewhat lower using ELR as compared to von Zeipel.  Models of the continuum polarization for a range of stellar types -- B0V, B1IV, B3V, and B8V -- are presented in Section 3, demonstrating that 1\% level polarizations are achievable at FUV wavelengths. Concluding remarks are given in Section 4.

\section{The Surface Properties of Rotating Stars}

To describe the properties of an axisymmetric atmosphere under rigid body rotation, we adopt Cartesian $(x_\ast,y_\ast,z_\ast)$, spherical $(r,\theta_\ast,\phi_\ast)$, and cylindrical $(a_\ast, \phi_\ast,z_\ast)$ coordinates.  Here $z_\ast$ is the spin axis, $\theta_\ast$ is co-latitude, and $\phi_\ast$ is azimuth.  The star spins with period $P$ and angular frequency $\Omega=2\pi / P$ about the $+z_\ast$ direction.

Our goal is to compute the continuum polarization for a star as a function of its rotation rate and viewing inclination for massive hot stars.  Doing so involves three primary considerations.  The first is to obtain the shape of the star since rotation will lead to a deviation from sphericity in the form of oblateness.  The second is to determine the temperature distribution across the star with co-latitude since the von Zeipel Theorem and related approaches predict gravity darkening at the equator. \cite{2025arXiv250920264W} find that the bolometric luminosity changes little (few percent) as a function of rotation. Consequently, temperature redistribution across the stellar surface will incorporate the assumption that
bolometric luminosity is conserved. Third is to predict the linear polarization across the star with latitude and longitude as projected onto the plane of the sky for a given observer inclination.

\subsection{Energy Transport in Rotating Stars}

We begin by reviewing the conditions of stellar equilibrium of the star, drawing on the works of \cite{1966MNRAS.132..201R, 1978trs..book.....T, 2009pfer.book.....M}.  The gravitational force $\vec{F_{\rm g}}$ is related to the gravitational potential $\Phi_g$ by 

\begin{equation}
\vec{F_g}~=~-\nabla \Phi_{\rm g}~,
\end{equation}

\noindent where $\nabla$ is the gradient operator. Under the assumption of rigid body rotation, the velocity at any point is $v=\Omega a_\ast$, where $a_\ast$ is the cylindrical radial
distance from the $z_\ast$ axis. 
The centrifugal force is $\vec{F_{\rm c}}=(v^2/a_\ast)\hat{a_\ast}$, that is

\begin{equation}
\vec{F_{\rm c}}~=~\Omega^2 a_\ast~\hat{a}_\ast~=~\Omega^2~(x_\ast^2 + y_\ast^2)^{1/2}~\hat{a}_\ast.
\end{equation}

\noindent The corresponding potential $\Phi_{\rm c}$ is

\begin{equation}
\Phi_{\rm c}~=~-\frac{1}{2}\Omega^2~(x_\ast^2 + y_\ast^2)~.
\end{equation}

\noindent The Laplacian of the total potential $\Phi$ is 

\begin{equation}
\nabla^2 \Phi~=~4\pi G\rho~-~2\Omega^2~.
\end{equation}

Using this in consideration of hydrostatic equilibrium, $\nabla P~=~-\rho~\nabla \Phi$, leads to a chain of argument in which each state variable of the gas is found to be constant on equipotentials, namely: $P=P(\Phi)$, $\rho~=~\rho(\Phi)$, and $T=T(\Phi)$.

The idea of ``gravity darkening'' arises from a consideration of radiative energy transport resulting from the preceding conclusions, with the radiative flux given by

\begin{equation}
\vec{{\cal F}}~=~-\frac{4c~a_{\rm rad}}{3}~\frac{T^3}{\kappa\rho}~\nabla T
\end{equation}

\noindent As shown in the previously cited works, a uniformly
rotating star cannot be in radiative thermal equilibrium which may lead
to slow currents within the star, known as Eddington-Sweet currents \citep{1986MNRAS.221...25M}. Such currents will not be very fast, so the equation of hydrostatic equilibrium remains valid. 

The gradient of the total potential $\Phi$ is set by the effective gravity $\vec{g}$, which includes the centrifugal term. ``Gravity darkening'' relates to how the effective gravity varies with latitude, being smaller at the equator.  That location will have a reduced flux and will thus be ``darker''. The scalar form of this relation is

\begin{equation}
{\cal F}~\propto~g~~~~~\Longrightarrow~~~~~~T_{\rm eff}~\propto~g^{1/4}
\label{eq:vonzT}
\end{equation}

\noindent and referred to as ``von Zeipel's law'' \citep{1924MNRAS..84..684V}. A rapidly rotating star will have a surface temperature that varies with latitude, highest at the pole and lowest at the equator.

\subsection{The Axisymmetric Shapes of Rotating Stars}

Most stars
are rather centrally condensed, so we can justify adopting a Roche model,
that is, a model with all the mass concentrated in the center. The local gravity is

\begin{equation}
\vec{g}~=~-\nabla\Phi~=~-\frac{G M}{r^2}~\hat{r}~+~\Omega^2 a_\ast~\hat{a}_\ast
\label{eq:effg}
\end{equation}

\noindent At the pole $a_\ast=0$ while at the equator $a_\ast=r$. We denote as the critical angular velocity for breakup, $\Omega_{\rm c}$,
such that $g=0$ at the equator. Letting $R_{\rm e}$ be the equatorial radius gives

\begin{equation}
g~=~0~=~-\frac{G M}{R_{\rm e}^2}~+~\Omega_c^2 R_{\rm e},
\end{equation}

\noindent resulting in

\begin{equation}
\Omega_{\rm c}^2~=~\frac{G M}{R_{\rm e}^3}~~~~~\Longrightarrow~~~~~~\Omega_c~=~\left(
                                                  \frac{GM}{R_{\rm e}^3}\right)^{1/2}
\end{equation}

\noindent At this critical condition, the equatorial radius will have a value of $R_{\rm e}~=~\frac{3}{2}~R_{\rm p}$. Additionally, the rotational velocity of the equator at break-up is

\begin{equation}
v_{\rm c}~=~R_{\rm e}~\Omega_{\rm c}~=~\left(\frac{GM}{R_{\rm e}}\right)^{1/2}~=~\left(\frac{2}{3}
          ~\frac{GM}{R_{\rm p}}\right)^{1/2} .
\end{equation}

Defining a scaled radius $x(\theta,W)=R(\theta,\Omega)/R_{\rm p}$, then the requirement that all points on the star lie on the same equipotential
surface as the pole (where $x=1$) leads to a cubic equation for $x$.
With $W=\Omega/\Omega_{\rm c}$ the angular velocity in terms of the critical velocity, the shape function has the solution:

\begin{eqnarray}
x(\theta_\ast,W) & = & \frac{3}{W~\sin\theta_\ast} \times \nonumber \\
 & & \cos\left\{\frac{1}{3}\left[\pi + \cos^{-1}
                  (W\sin\theta_\ast)\right]\right\}
                  \label{eq:rochex}
\end{eqnarray}

\begin{figure}
\includegraphics[width=\columnwidth]{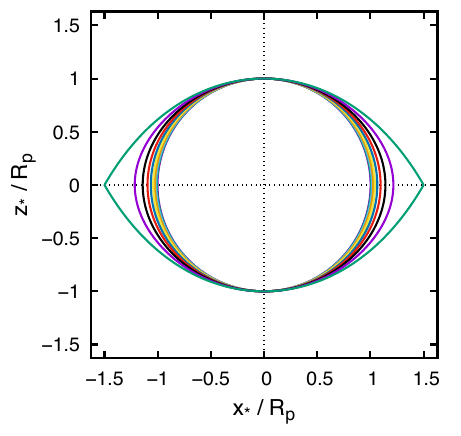} 
\caption{The shapes of rapidly rotating stars with $W$ ranging from 0.0 (non-rotating and circular) to 1.0 (largest equatorial extent) in steps of 0.1.  The coordinates are normalized to the polar radius which is assumed not to change with rotation.
\label{fig:surface}}
\end{figure}

Figure~\ref{fig:surface} shows the shapes obtained from this equation for various values of $W$, from non-rotating ($W=0$ for a spherical star that is circular in cross-section) to critical ($W=1$ displaying the greatest equatorial extent).
Defining the critical rate $\Omega_{\rm c}^2=(2/3)^3~GM/R_{\rm p}^3$,
and using $\Omega^2/\Omega_{\rm c}^2=W^2$, we obtain finally

\begin{equation}
g = \left(\frac{GM}{R_{\rm p}^2}\right)~\frac{1}{x^2}\left\{\cos^2\theta_\ast +
 \sin^2\theta_\ast\left[1-\frac{8}{27}x^3W^2\right]^2\right\}^{1/2}.
\end{equation}

\noindent This magnitude of the local surface gravity is largest at the pole and decreases monotonically with latitude to its smallest value at the equator.

\begin{figure*}
\includegraphics[width=\textwidth]{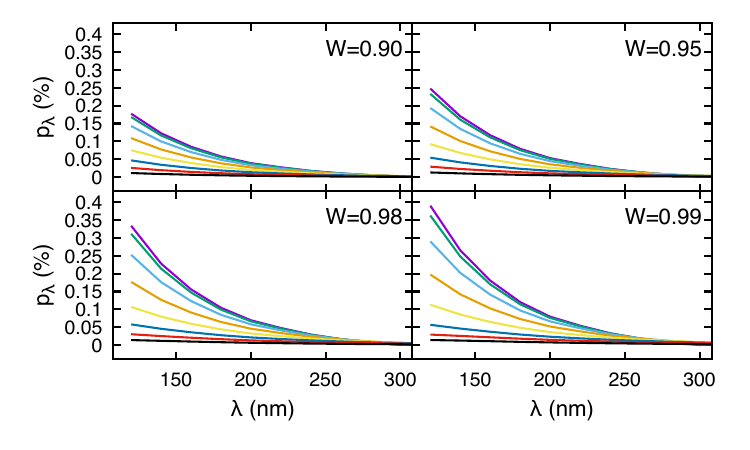} 
\caption{Isolating the effect of rotational distortion for stellar polarization.  These models assume the temperature and gravity are the same as the non-rotating B1IV star model everywhere but now for non-spherical shape from rotational distortion.  The polarization contributions are actually negative at short wavelengths with $q<0$ but plotted here as positive definite polarization $p_\lambda = |q|$.  Each panel is for the rotation rate $W$ as labeled. Colors are for inclinations of $20^\circ-90^\circ$ in $10^\circ$ increments, with edge-on giving the highest polarization.  As will be seen, for this spectral class, the geometrical distortion accounts for roughly half of the polarization signal.
\label{fig:oblate}}
\end{figure*}

Our methodology for determining continuum polarization calculations will be described more fully in Section 3.  However, it is useful here to isolate the effect of the non-spherical shape for producing a polarimetric signal.  Figure~\ref{fig:oblate} displays continuum polarizations for rotationally distorted atmospheres described by the Roche model.  The four panels are for rotation rates as labeled using the B1IV model star of Table~\ref{tab:stars}.  The curves in each panel are for different viewing inclinations from $20^\circ-90^\circ$, with edge-on producing the highest polarization curve.  We show wavelengths from FUV to NUV wavelengths.  The figure reveals how much polarization is derived purely from the rotationally distorted shape.  As will be seen in Section 3, the polarizations arising from rotational distortion account for nearly half the amplitude when the latitude-dependent temperature and gravity are included.

\subsection{The Relative Temperature Distributions for Rotating Stars}

Our objective is to compare the polarizations between the von Zeipel and ELR approaches. To do so, we need a choice for how to scale temperature with latitude.
This is not a significant problem, because the nature of our study is to make comparisons between stars, and between places on stars, where some other property is kept the same when rotation is added.  Whatever other property is kept the same is somewhat arbitrary. If for example we kept stellar attributes like mass and age the same, we would not necessarily know they were the same when we compare against observed stars. We seek to establish a property that does not change with rotation.
For example, we might keep the polar temperature fixed regardless of $W$. 
However, fixing the polar temperature means that with less flux from the equatorial zone, despite increased surface area, the luminosity will be lower than its static counterpart.  
Instead, we prefer to keep the stellar luminosity the same for all $W$, which results in an increase of polar temperature.

Since we have made the choice to compare stars with the same luminosity as we assess the observable effects of rotation, we find it necessary to correct this luminosity shortfall by scaling up the absolute temperatures.
Our primary goal is to make a comparison of the degree of polarization, further deemphasizing the importance of absolute temperatures, since relative temperatures are of greater impact on polarization due to the high degree of cancellation involved.

 On this point, we recognize that if all stellar properties except rotation are held fixed, introducing rotation can often reduce the stellar luminosity.  The virial theorem can be invoked to understand this, because that theorem requires the added rotational kinetic energy to be matched by twice that amount of added gravitational energy. This indicates a contraction of the stellar interior that harbors the radiatively diffusing luminosity, hence lowering the luminosity supply given the fairly fixed temperatures set by fusion.
However, the rotational kinetic energy of even a critically rotating star will represent only a few percent of the virialized energy\footnote{This conclusion is consistent with more detailed calculations appearing in Tab.~6.1 of \cite{2025arXiv250920264W}.}, so the distinction is hardly crucial for the comparisons we are making.
Moreover, it is not in practice possible to make observations of two stars with all attributes but rotation the same, as those attributes would not even be known until the rotation has been assessed.
We therefore regard our comparisons of differently rotating stars at matched luminosity to be an arbitrary choice that should have little impact on the conclusions, given the relatively minor effects of rotation on luminosity.

\subsubsection{The Case of von Zeipel}

To conserve luminosity requires calculation of the stellar surface area. We seek the surface unit normal $\hat{n}$ as a function co-latitude.  The local normal makes an angle
$\delta$ with the $z_\ast$-axis. To determine its value, we consider the local vector gravity, since $\hat{n}=-\hat{g}$.  The expression is

\begin{equation}
\vec{g}=\left[\frac{-g_{\rm p}}{x^2}+\Omega^2~x~R_{\rm p}\right]\sin\theta_\ast~\hat{a}_\ast
     +\left[\frac{-g_{\rm p}}{x^2}\right]\cos\theta_\ast~\hat{z}_\ast,
\end{equation}

\noindent where $g_{\rm p}=GM/{R_p}^2~$ is the magnitude of the gravity at the pole. The ratio of the gravitational components relates to $\delta$ through 

\begin{equation}
\tan\delta~=~\left\{1~-~\frac{\Omega^2 x^3 R_{\rm p}}{g_{\rm p}}\right\}~\tan\theta_\ast.
\end{equation}

\noindent And the last term in the braces simplifies to 

\begin{equation}
\frac{\Omega^2 x^3 R_{\rm p}}{g_{\rm p}}=\Omega^2\frac{R_{\rm p}^3 x^3}{GM}=\frac{8}{27} x^3 \frac{\Omega^2}
  {\Omega_c^2}=\frac{8}{27} x^3 W^2,
\end{equation}

\noindent giving

\begin{equation}
\tan\delta~=~\left\{1~-~\frac{8}{27} x^3 W^2\right\}~\tan\theta_\ast~~.
\end{equation}

The surface area of the star derives from integration over the range $0\le \phi_\ast \le 2\pi$ and $0\le \theta_\ast \le \pi$. The differential element of area in
the radial direction $\hat{r}$ is 

\begin{eqnarray}
dA'~& =& ~{R^2(\theta_\ast,W)}~d\phi~\sin\theta_\ast~d\theta_\ast~ \\
 & = &~R_{\rm p}^2~~{x^2(\theta_\ast,W)}~d\phi_\ast 
   ~\sin\theta_\ast~d\theta_\ast
\end{eqnarray}

\noindent with $x(\theta_\ast,W)$ from equation~(\ref{eq:rochex}). The stellar surface element $dA$ is inclined to the radial 
surface element $dA'$ by the angle $\theta-\delta$, so $dA$ projects onto $dA'$ by
a factor of $\hat{n}\cdot\hat{r} = \cos(\theta-\delta)$. Thus an expression for the stellar surface area is

\begin{equation}
A_\ast~=~4\pi~R_{\rm p}^2~~~\frac{1}{2}\int_0^\pi {x^2(\theta_\ast,W)}~\frac{\sin\theta_\ast}
                    {\cos(\theta_\ast-\delta)}~d\theta_\ast 
\end{equation}

\noindent Table~\ref{tab:temps} gives example values for how the stellar surface area gorws with $W$. 


The luminosity of the rotating star will be given by the flux from each surface
element integrated over the star. Assigning the temperature at the
pole, the bolometric stellar luminosity becomes

\begin{eqnarray}
L_\ast & = & 2\pi R_{\rm p}^2~\sigma T_{\rm p}^4 \times \nonumber \\
 & & \int_0^\pi {x(\theta_\ast,W)}^2
  \frac{\sin\theta_\ast}{\cos(\theta_\ast-\delta)}\left[\frac{T(\theta_\ast)}{T_{\rm p}}\right]^4
  d\theta_\ast.
\end{eqnarray}

\noindent Substitution into the integrand for luminosity gives 

\begin{eqnarray}
\left[\frac{T(\theta_\ast)}{T_{\rm p}}\right]^4 & = & \frac{g(\theta_\ast)}{g_{\rm p}}~=~
\frac{1}{x^2}~\left\{
\cos^2\theta_\ast \right. \\ \nonumber
& & \left. + \sin^2\theta_\ast\left[1~-~\frac{8}{27}~x^3~W^2\right]^2\right\}^{1/2}~~.
\end{eqnarray}

\noindent After simplifying, 

\begin{eqnarray}
L_\ast & = & 4\pi~R_{\rm p}^2~\sigma T_{\rm p}^4~~~\frac{1}{2}\int_0^\pi \frac{\sin\theta_\ast}
  {\cos(\theta_\ast-\delta)} \times \\ \nonumber
  & & \left\{\cos^2\theta_\ast+ 
\sin^2\theta_\ast\left[1~-~\frac{8}{27}~x^3~W^2\right]^2\right\}^{1/2}d\theta_\ast
\end{eqnarray}

\noindent The leading term, $4\pi~R_{\rm p}^2~\sigma T_{\rm p}^4$, is just the luminosity of a
spherical star of radius $R_{\rm p}$ and surface temperature $T_{\rm p}$. The remaining
integral gives the ratio of the luminosity of a rotating star with polar
radius $R_{\rm p}$ to a non-rotating star of radius $R_{\rm p}$ and surface temperature 
$T_{\rm p}$. Some results are given in Table~\ref{tab:temps}. We see that the total
luminosity of the rotating stars will be substantially less than the non-rotating
counterpart if the polar temperature is held to that of the non-rotating star.


Increasing the temperature at
all $\theta$ by the factor $(L_{\rm sph}/L)^{1/4}$ ensures constant total luminosity  at all rotations. This factor, which is just $T_{\rm p}/T_{\rm sph}$, is given in the last 
column of Table~\ref{tab:temps} where at critical rotation the pole is nearly 12\% hotter than with no rotation.
Thus we see that the relative differences in temperature from place to place on the star can be significantly affected by rotation, and this is the primary effect we wish to track as we explore the observable polarization, in addition to the effect on the \textit{shape} of the surface.

\begin{table}
\caption{Stellar Distortion Contrasts with $W$   \label{tab:temps}}
\begin{tabular}{cccc}
\hline\hline {$W=\Omega/\Omega_{\rm c}$}&{$A/A_{\rm sph}$}&{$L/L_{\rm sph}$}&{$T_{\rm p}/
                    T_{\rm sph}$}\\[0.5ex]
\hline
0.5 & 1.056 & 0.9465   & 1.0139 \\
0.8 & 1.186 & 0.8375   & 1.0453 \\
0.9 & 1.282 & 0.7705   & 1.0674 \\
0.95 & 1.363 & 0.7222  & 1.0848 \\
0.99 & 1.485 & 0.6645  & 1.1076 \\
1.0 &  1.577 & 0.6394  & 1.1183 \\
\hline
\end{tabular}
\end{table}

\begin{figure}
\includegraphics[width=\columnwidth]{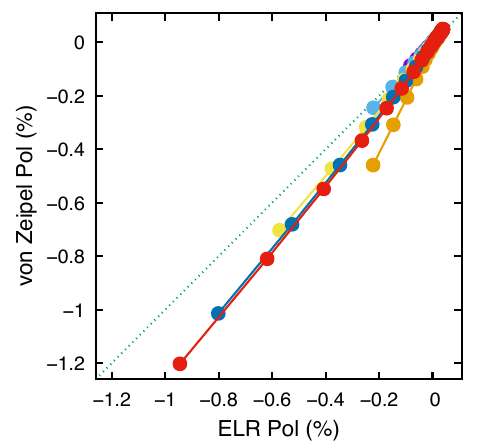} 
\caption{Compares polarization using von Zeipel against ELR.  The diagonal dashed line is where the two approaches produce the same polarization.  For a nonrotating star of type B1IV, the different colors correspond to rotations rates of $\Omega=0.60$, 0.80, 0.90, 0.95, 0.98, and 0.99 in order of increasing polarization.  The circles are for different wavelengths, starting from 120 nm for the greatest negative polarizations, and  declining toward optical wavelengths at the upper right.
\label{fig:vonz}}
\end{figure}

\subsubsection{The Case of Espinosa Lara and Rieutord (ELR)}

As noted earlier, the von Zeipel law neglects effects of the  
Eddington-Sweet currents. Observations of nearby rapid rotators with stellar 
interferometers have found that the von Zeipel law overestimates the temperature 
contrast between the pole and equator. 
In light of these results, ELR presented a method
of computing an improved gravity darkening law based on the assumption that the flux
throughout the star is antiparallel to the effective gravity. They show that the 
the darkening law can be obtained by solving one trancendental equation. 

ELR introduce the following parameter,

\begin{equation}
\tau~=~\frac{1}{3}\tilde{W}^2\tilde{r}^3 \cos^3\theta_\ast~+~\cos\theta_\ast~+~\ln\left(\tan
           \frac{\theta_\ast}{2}\right)
\end{equation}

\noindent where $\theta$ is the co-latitude and $\tilde{r}= R(\theta_\ast,\Omega)/R_{\rm e}(\Omega)$. ELR assume 
that the star is centrally condensed so that the shape will be given by the Roche
model. The factor $\tilde{W}$ is a rotation rate but relative to the Keplerian angular velocity at the star's equator.  So 

\begin{equation}
    \tilde{W} = \Omega/\Omega_{\rm kep}.
\end{equation}

\noindent with $\Omega_{\rm kep}=(GM/R_{\rm e}^3)
^{1/2}$.  In terms of $x(\theta_\ast,W)$, $\tilde{r}= x(\theta_\ast,W)/x(\pi/2,W)$.  As a result:

\begin{equation}
\tilde{W}~=~{\left[\frac{2}{3}~x\left (W,\frac{\pi}{2}\right )\right]}^{3/2}~W.
\end{equation}

ELR introduce a new angular parameter $\Theta$ that is defined as a function of $\tau$ by the transcendental equation:

\begin{equation}
\tau~=~\cos\Theta~+~\ln\left(\tan\frac{\Theta}{2}\right)
\end{equation}

\noindent Then they show that the deviation from the von Zeipel law is given by the quantity

\begin{equation}
F(\tilde{\Omega},\theta_\ast)~=~\frac{\tan^2 \Theta}{\tan^2 \theta_\ast}
\end{equation}

\noindent such that

\begin{equation}
T_{\rm eff}(\theta) 
   ~\propto~\left[F_\Omega(\theta)~g(\theta)\right]^{1/4}.
\end{equation}

\noindent While the results for this formulation approaches the von Zeipel law for
low angular velocities ($W < 0.5$), it differs substantially from von Zeipel
for high angular velocities. For example, for $W=0.6,0.8,0.9,0.95$ and 0.99
($\tilde{W}$ = 0.359, 0.531, 0.657, 0.750 and 0.883), the $T_{\rm e}/T_{\rm p}$ ratios
from the ELR formulation are, respectively,
0.941, 0.883, 0.832, 0.789 and 0.713.
For the Von Zeipel law, the corresponding $T_{\rm e}/T_{\rm p}$ ratios are 0.937, 0.862, 0.788, 0.719 and 0.581. Bear in mind that the total
radiated flux goes as $T^4$, so these differences are magnified in the flux contrasts.


\section{Continuum Polarization from Rotating Stars}

In stellar atmospheres where the scattering is a substantial fraction of 
the opacity, the emergent radiation will be partially polarized. For isolated
spherical stars that are spatially unresolved, there will be zero net 
polarization owing to cancellation when integrating across the face of the 
star. Only by breaking spherical symmetry can we have observable net 
polarization. Rotation is one way to break that symmetry. Even at low rates
of rotation, symmetry breaking can occur across spectral lines, the so-called
Ohman effect \citep[e.g.,][]{harrington25}. This arises over a narrow 
frequency range when an absorption line reduces the flux from part of the 
star, but, due to the rotational Doppler shift, flux from other parts of the 
stellar surface are unaffected. However, at adjacent frequencies, other parts
of the stellar surface will be blocked, resulting in polarization at a 
contrasting angle which, when integrated over a frequency range exceeding the
maximum Doppler shift, will result in net zero polarization. Thus measurements
at low spectral resolution, to the extent that a slowly rotating star remains 
very nearly spherical, will show no net polarization. For a net continuum 
polarization to arise requires rapid rotation in excess of $W \sim 0.5$, where 
distortion in the shape of the star becomes perceptable, and, simultaneously,
gravity darkening effects appear.

To compute the continuum polarization, we follow the methods outlined in
\cite{harrington25}. In summary, TLUSTY model atmospheres for a grid of
$T_{\rm eff}$ and $\log g$ were used. The opacities and scattering with depth were
extracted, and calculations for the emergent Stokes I and Q intensities as
a function of angle with respect to the normal to the atmosphere's surface were made.
Then for a given rotation rate $W$ and spectral class for a non-rotating star,
we use the considerations of the preceding sections to obtain the 
non-spherical surface shape along with gravity and temperature distributions with latitude
using the ELR approach for temperature. The stellar surface is modeled by patches of 
area, each patch one degree in latitude and longitude. 
For each patch, we compute the direction cosine $\mu$ toward the observer, excluding all patches not observable ($\mu < 0$).  

\begin{table}
\begin{center}
\caption{Stellar Properties \label{tab:stars}}
\begin{tabular}{rl}
\hline\hline Property & Value \\ \hline 
B0V with $\Omega=0$: & L\&H$^\dag$ \\
$T_{\rm eff}$ (K) & 29,000 \\
$\log g$ (cm/s$^2$) & 3.9 \\
$M_\ast~(M_\odot)$ & 17.5 \\
$R_\ast~(R_\odot)$ & 7.4 \\
$L_\ast~(L_\odot)$ & 35,000 \\ 
 & \\
 B1IV with $\Omega=0$: & SK$^\dag$ \\
$T_{\rm eff}$ (K) & 25,000 \\
$\log g$ (cm/s$^2$) & 3.6 \\
$M_\ast~(M_\odot)$ & 16.3 \\
$R_\ast~(R_\odot)$ & 10.1 \\
$L_\ast~(L_\odot)$ & 34,000 \\ 
 & \\
B3V with $\Omega=0$: & L\&H$^\dag$ \\
$T_{\rm eff}$ (K) & 18,000 \\
$\log g$ (cm/s$^2$) & 4.0 \\
$M_\ast~(M_\odot)$ & 8.0 \\
$R_\ast~(R_\odot)$ & 4.7 \\
$L_\ast~(L_\odot)$ & 2,100 \\ 
 & \\
B8V with $\Omega=0$: & SK$^\dag$ \\
$T_{\rm eff}$ (K) & 12,000 \\
$\log g$ (cm/s$^2$) & 4.00 \\
$M_\ast~(M_\odot)$ & 3.8 \\
$R_\ast~(R_\odot)$ & 3.1 \\
$L_\ast~(L_\odot)$ & 200 \\\hline
\end{tabular}
\end{center}

{\small $^\dag$ Stellar parameters from L\&H for \cite{2007ApJS..169...83L} and SK for Schmidt-Kaler in \cite{1982lbg6.conf.....A}.}
\end{table}

The classical Be stars are preferentially found at earlier spectral types,
generally main sequence or subgiants. To illustrate continuum polarization
levels, we selected 3 non-rotating star types given in Table~\ref{tab:stars}: B0V, B1IV, 
and B3V. For each case, the relevant $T_{\rm eff}$ and $\log g$ were obtained at each latitude by interpolation.

The Bn stars are more common among cooler B stars. We thus include a B8V star in our
study. The problem here is that the rapidly rotating ($W = 0.95$ or 0.98) B8V
star has an equatorial temperature that is so low ($\sim 10^4$~K) that the 
TLUSTY models were not available. We therefore used the Kurucz
models \citep[e.g.,][]{2003IAUS..210P.A20C} as input to the synspec205 program \citep{hubeny2011}, modified to output the
needed depth-dependent scattering and opacities, for our computation of the
emergent polarized radiation.

The specific Stokes I and Q luminosities as functions of viewing 
inclination are given by:

\begin{equation}
L_I = 8\pi\, R_{\rm p}^2\int_{-\pi}^{+\pi} \int_0^\pi \mu I_\lambda(\theta_\ast,\mu)~
  \frac{x^2 \sin\theta_\ast}{\cos(\theta_\ast-\delta)}~d\phi_\ast~d\theta_\ast,
\end{equation}

\noindent and

\begin{eqnarray}
L_Q & = & 8 \pi \, R_{\rm p}^2\int_{-\pi}^{+\pi} \int_0^\pi \mu Q_\lambda(\theta_\ast,\mu)~\cos(2\xi)~\nonumber \\ 
  & &  \frac{x^2 \sin\theta_\ast}{\cos(\theta_\ast-\delta)}~d\phi_\ast~d\theta_\ast.
\end{eqnarray}

\noindent Here, the $I_\lambda(\mu)$ and $Q_\lambda(\mu)$ are from the radiative transfer solutions with $\theta_\ast$ identifying the co-latitude corresponding to the stellar atmosphere model in $g$ and $T$.  They include the variation of temperature and gravity with co-latitude on the star.  The direction cosine $\mu = \hat{n} \cdot \hat{z}$, for $\hat{n}$ the local normal to the surface and $\hat{z}$ the unit vector toward the observer.  The integration proceeds over the entire surface of the star and simply ignores contributions whenever $\mu<0$, which signifies an areal surface patch that is occulted (i.e., behind the star with respect to the observer).  The coordinate $\xi$ is an azimuthal angle about the observer sightline.

\begin{figure*}
\includegraphics[width=\textwidth]{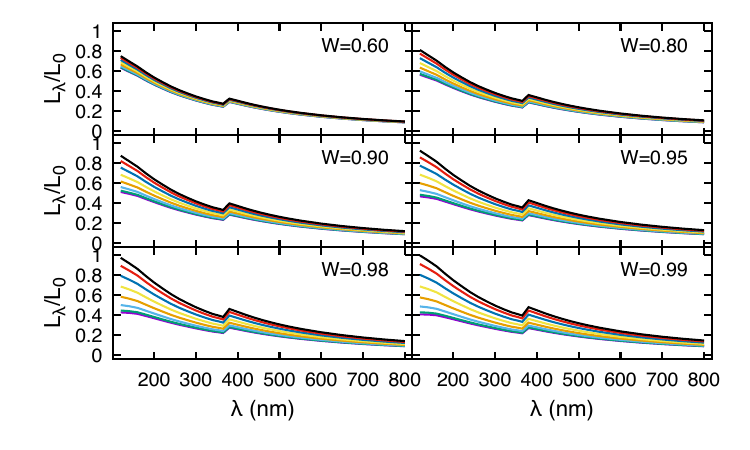} 
\caption{Luminosity distributions with $L_\lambda = L_I$ for rotation $W$ and viewing inclination $i$ for a B1IV star.  Each panel is labeled for $W$.  Colors are for inclinations as in Fig.~\ref{fig:oblate}.  The spectra have been scaled by the brightest value of all models, and each panel has the same vertical axis scale.
\label{fig:B1IVflx}}
\end{figure*}


While our models employ the results from the ELR approach, it is useful compare those with the von Zeipel approximation.
Figure~\ref{fig:vonz} shows how the continuum polarization, $100\% \times L_Q/L_I$, differs between these two predictions, using the B1IV star seen from the equator. The figures shows polarization as negative, indicating a polarization position angle that is orthogonal to the spin axis.  The dotted line signifies equal polarization in both von Zeipel and ELR approaches.  The dots are wavelengths from 120~nm (highest polarization for each color) to 800~nm.  The six colors are for different rotation rates $W=0.6, 0.8, 0.9, 0.95, 0.98,$ and 0.99 (the red curve), with faster rotation resulting in higher (negative) polarization.  

The fact that the curves overlap significantly is not important.  What is important is that all the points lie below the diagonal, although not too far away.  The ELR model consistently produces lower polarizations than does von Zeipel, but the reduction is of order unity.  In the figures that follow, we employ ELR exclusively.

Figure \ref{fig:B1IVflx} displays the specrtral energy distributions from 120 nm to 800 nm.  The six panels are for the rotation rates $W$ as labeled.  Like Figure~\ref{fig:surface}), the different colored curves are for viewing inclinations $i=20^\circ-90^\circ$ in $10^\circ$ increments.  
The purple curve is for edge-on.  



\begin{figure}
\includegraphics[width=\columnwidth]{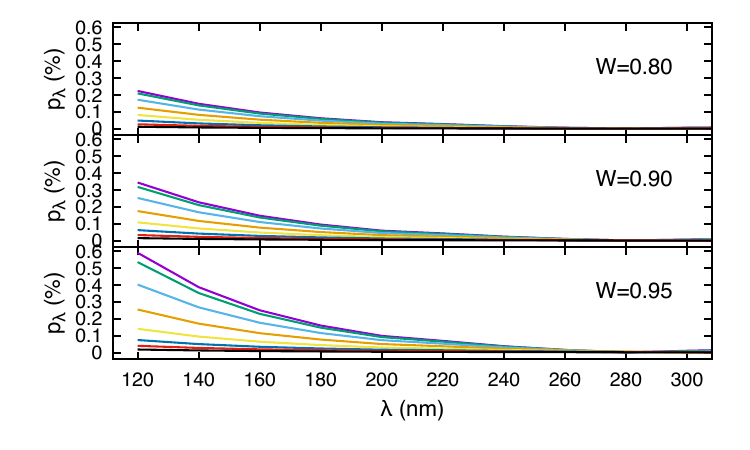} 
\caption{Polarization for the B1IV star model, highlighting shorter wavelengths below 300~nm, with $\Omega=0.80$, 0.90, and 0.95 as labeled.  Colors are for inclinations as in Figs.~\ref{fig:oblate} and \ref{fig:B1IVflx}.
\label{fig:B1IVzoom}}
\end{figure}

\begin{figure}
\includegraphics[width=\columnwidth]{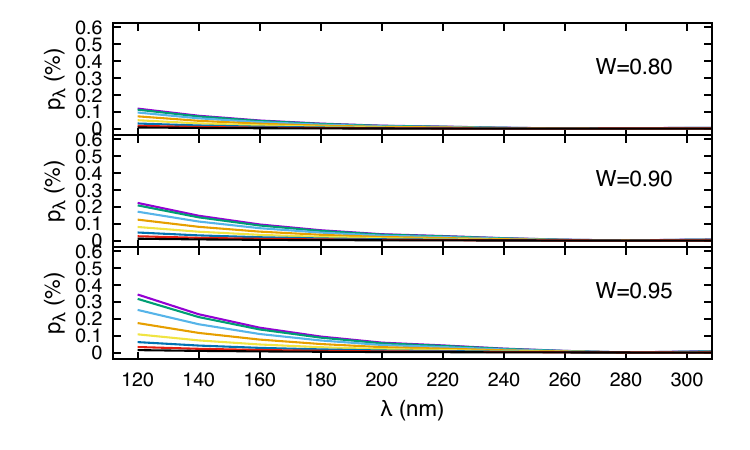} 
\caption{As in Fig.~\ref{fig:B1IVzoom} now for a hotter yet smaller B0V star.
\label{fig:B0Vzoom}}
\end{figure}

\begin{figure}
\includegraphics[width=\columnwidth]{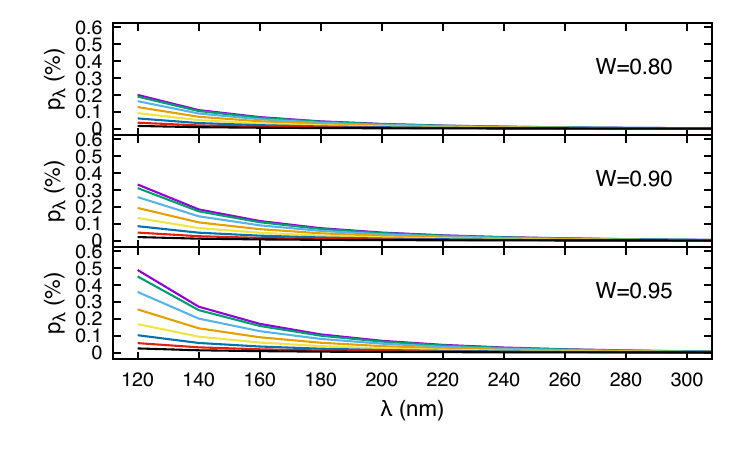} 
\caption{As in Fig.~\ref{fig:B0Vzoom} now for a cooler and smaller B3V star.
\label{fig:B3Vzoom}}
\end{figure}

\begin{figure}
\includegraphics[width=\columnwidth]{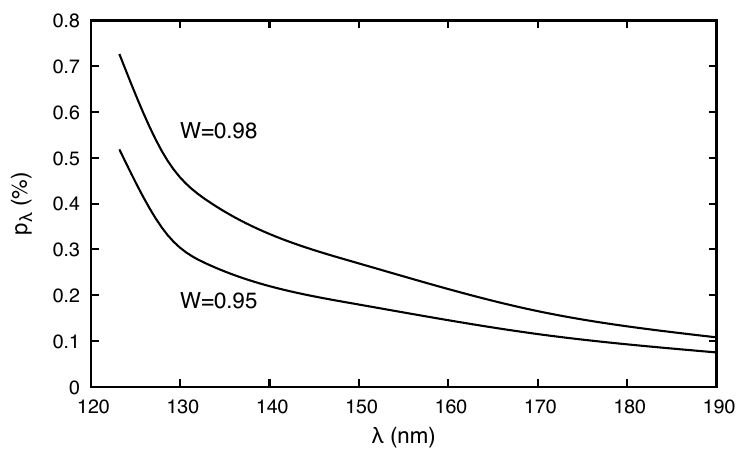} 
\caption{Polarization spectrum for a B8V star showing only $W$ of 0.95 and 0.98, as labeled, and only for an edge-on inclination of $i=90^\circ$. Note the scale is up to 0.8\%, whereas Figs.~\ref{fig:B1IVzoom}--\ref{fig:B3Vzoom} only go to 0.6\%. 
\label{fig:B8Vzoom}}
\end{figure}

Ignoring the optical wavelengths, Figures~\ref{fig:B1IVzoom}--\ref{fig:B8Vzoom} zoom into the polarization distributions from 120 nm to about 300 nm.  Figures~\ref{fig:B1IVzoom}, \ref{fig:B0Vzoom}, and \ref{fig:B3Vzoom} are respectively for stars that when non-rotating are the B1IV, B0V, and B3V spectral types (again, see Tab.~\ref{tab:stars}).  Now three panels are shown with $W=0.80$ (top), 0.90 (middle), and 0.95 (bottom). Figure~\ref{fig:B8Vzoom} is for a B8V star when non-rotating. Rotations of $W=0.95$ and 0.98 are shown, both for $i=90^\circ$, and the longest wavelength is only 190~nm to highlight the trend at the shortest wavelengths shown.

The optical is not shown because the polarization is so low at those wavelengths, as highlighted by \cite{harrington25} in their study of the \"{O}hman effect.  Although they considered spherical atmospheres, examples of limb polarization for different star types were shown.  All had very low polarizations at optical wavelengths.  Clearly breaking symmetry hardly matters for modifying the polarimetric cancellation at wavelengths where there is little amplitude to begin with.  At issue is that for hot stars, the optical wavelengths are always in the Rayleigh-Jeans tail, and limb polarization is small.  Although extreme rotation rates can drive equatorial temperatures to low values moving the Wien peak into the optical, those regions have so little flux, the polarization remains low.  It is at short UV and especially FUV wavelengths where the polarization rises to significant levels, corresponding to be near or short of the Wien peak where the source function gradient is steep.  

Returning to the FUV-NUV wavelengths in the 4 figures, the weakest polarization is for the B0V star.  This is because the shortest wavelengths plotted are just longward of the Wien peak.  Additionally, being main sequence, the gravities are higher which suppresses the polarization level.  The B3V star has relatively somewhat higher polarization.  Still main sequence with high gravity, the lower temperature moves the Wien peak squarely into the UV band.  The highest polarizations are for the B1IV star.  A little cooler than the B0V star and a little lower gravity than the B3V star. At $i=90^\circ$ B8V polarization at $W=0.95$ is similar overall to the B3V star. A faster rotation of $W=0.98$ is shown where a maximum polarization of nearly 0.75\% is achieved at the shortest wavelength.

At 95\% of critical, all the models show polarization near half a percent at the shortest wavelength of 120 nm for an edge-on inclination.  By 200 nm, the polarization drops to about 0.1\%, again for edge-on.  Even with $W=0.8$, polarization of a 0.1\% can be obtained around 120~nm, for high inclinations.

Except for large ground-based telescopes, the best means for inferring rapid rotation from massive stars is through FUV spectropolarimetry.  As noted by \cite{2025Ap&SS.370...57I}, the polarization from rotationally distorted hot star atmospheres is rising where the interstellar polarization is dropping.  For hot stars the circumstellar polarization likely arises from Thomson scattering.  Except for optically thick winds, like the Wolf-Rayet stars, the chromatic signature of Thomson scattering is mainly a flat polarization, except when other absorptive opacity sources become relevant \citep[again, see][]{2025Ap&SS.370...57I}.  Hence the polarimetric signature for fast rotation is a rising polarization toward shorter FUV wavelengths -- distinctive from electron scattering and showing the opposite trend expected from interstellar polarization.

\section{Conclusions}

The internal physics of rapidly rotating B stars remains an unsolved question in stellar structure and evolution \citep[e.g.,][]{2024IAUS..361..343H}. An important observable aspect of that internal physics is the equatorial oblation and consequent ``gravity darkening'' that is visible at the surface, which yields net photospheric linear polarization by reducing the global cancellation of that polarization \citep{1968ApJ...151.1051H}.  Because this polarization is null in pure spherical symmetry, any detectable signal is an indicator of the effects of rapid rotation and is influenced by how that rotation is altering the stellar interior.

We have modeled that polarization signal in the FUV band where the B star continuum peaks.  This is where the polarization is strongest, both because that is where the Wien tail of the Planck function maximizes the brightness contrasts stemming from pole-to-equator temperature differences, and because polarization-reducing absorbing opacity is reduced in the FUV, at wavelengths well separated from the Balmer continuum edge.

We find that the observable FUV continuum polarization, which is enhanced whenever global cancellation is suppressed, stems about equally from the distortion of the stellar shape, and from the gravity darkening of the equator.  The gravity darkening is slightly more important if simple von Zeipel gravity darkening is assumed, but if the more physically motivated ELR model is used, then equatorial oblation is the slightly more important effect.  We also find that using ELR reduces the polarization by a fairly ubiquitous 3/4 factor relative to von Zeipel, which is enough to help distinguish the two models, but not so much as to render the polarization unobservable in the ELR treatment.

We find general agreement with the results of \cite{1991ApJS...77..541C}, though our polarizations are fairly consistently about 10\% less than theirs, for our nominal results that assume ELR.  Since \cite{1991ApJS...77..541C} used the more pronounced gravity darkening of von Zeipel, the fact that our polarizations are only 10\% weaker, not 25\% weaker as for the abovementioned 3/4 reduction factor from ELR implies we obtain a broadly 15\% increase stemming from some aspect of the updated atmospheres we are using.  All told, this implies the conclusions from \cite{1991ApJS...77..541C} about the overall observability of FUV continuum polarization in rapidly rotating early B stars continue to hold to a good approximation in the modern scenario.

A related conclusion of this work is we also concur with past findings in general \cite{1991ApJS...77..541C}, and for Regulus in particular \cite{2022Ap&SS.367..124J}, that the polarization near Lyman $\alpha$ is canonically about double the polarization at 150~nm, over a broad range of models.  Hence, the ability for a spaceborne instrument to cover wavelengths right down to the 122~nm limit of magnesium fluoride, such as for the {\em Polstar} spectropolarimetry concept \citep{2025Galax..13...40I} and other potential FUV missions, is important for maximizing the detectable polarization signal.  Furthermore, the presence of this steep rise toward the FUV is a telltale indicator of rapid photospheric rotation, readily distinguished from other types of polarization \citep[like winds, circumstellar disks, and the interstellar medium][]{2025Galax..13...40I} that do not show so steep a rise there. 
Quasiperiodic modulation in the time domain of this steeply rising signal could also help single out rotating surface features like spots, though only if the spots are large enough \citep[e.g.,][]{2024MNRAS.529..374B}.

Since this steep rise toward FUV wavelengths is an unambiguous diagnostic of rapid rotation, it behooves us to understand its origin.
It is even steeper than can be explained strictly from the behavior of the Wien approximation to the FUV Planck function (for which brightness contrast ratios at fixed $\lambda$ depend strictly on $\lambda T$), so there must be additional contribution from explicitly wavelength and temperature dependent effects that involve the opacity structure.  Although this has yet to be demonstrated quantitatively, these opacity effects might center on the relatively weaker absorption toward FUV wavelengths, as compared to the ubiquitous Thomson scattering opacity. 

We also concur with the finding \cite{1991ApJS...77..541C} that lower surface gravity at fixed spectral type produces significant polarization increases, making polarization stronger and hence easier to detect in rapidly rotating evolved stars such as $\epsilon$ Ori \cite{2025arXiv250702276O}.  At fixed $\lambda T$ Planck contrasts are kept constant, yet a polarization increase is still seen for low gravity, so it must arise from a reduction in absorption as expected from the implied lower densities.

It should be noted that this paper focuses primarily on early B stars, which when sufficiently rapidly rotating, often produce orbiting disks, making them classical Be stars \citep{2013A&ARv..21...69R}.   Since disks will produce their own polarization in the optical and FUV \citep[e.g.,][]{2005ASPC..343..270B}, those polarimetric signals will be superimposed on the photospheric ones we derive here.  Again the telltale steep rise toward the Lyman limit will prove helpful in distinguishing the effects of photospheric rapid rotation from disk-like polarization contributions. One might also expect the photospheric signal to be more consistent with time, whereas disk emission can be highly variable over months to years \citep[e.g.,][]{2019MNRAS.486.5139P}.  

As disk polarization simulations become more adept at treating metal-line opacity, we will also include the combined polarization of the photosphere and a variable disk, to address FUV polarization in classical Be stars. In the future, we also plan to extend this study to cooler B stars that include a richer population of diskless Bn stars.  This continuing investigation pathway is ultimately intended to support the interpretation of the steeply rising FUV continuum polarization from rapidly rotating B stars of all types, to constrain better their angular momentum distribution and evolution.


\section*{Funding Statement}

The authors declare that no funds, grants, or other support were received during the preparation of this manuscript.

\section*{Ethics Approval}

Not applicable

\bibliography{gdark}

\end{document}